\documentclass[english,twocolumn,showpacs,preprintnumbers,amsmath,amssymb,prb]{revtex4-1}

\usepackage{graphicx}
\usepackage{color}

\usepackage{epsfig}

\makeatother

\begin{document}

\title{Phonon dispersion in two-dimensional solids from atomic probability
distributions }

\author{R. Ram\'{\i}rez\footnote{Electronic mail:ramirez@icmm.csic.es} and
C. P. Herrero}

\affiliation{Instituto de Ciencia de Materiales de Madrid (ICMM), Consejo Superior
de Investigaciones Cient\'{\i}ficas (CSIC), Campus de Cantoblanco,
28049 Madrid, Spain }

\begin{abstract}
We propose a harmonic linear response (HLR) method to calculate the
phonon dispersion relations of two-dimensional (2D) layers from equilibrium
simulations at finite temperature. This HLR approach is based on the
linear response of the system, as derived from the analysis of its
centroid density in equilibrium path integral simulations. In the
classical limit, this approach is closely related to those methods
that study vibrational properties by the diagonalization of the covariance
matrix of atomic fluctuations. The validity of the method is tested
in the calculation of the phonon dispersion relations of a graphene
monolayer, a graphene bilayer, and graphane. Anharmonic effects in
the phonon dispersion relations of graphene are demonstrated by the
calculation of the temperature dependence of the following observables:
the kinetic energy of the carbon atoms, the vibrational frequency
of the optical $E_{2g}$ mode, and the elastic moduli of the layer.
\end{abstract}

\maketitle

\section{Introduction}

The analysis of vibrational modes of solids and molecules from equilibrium
simulations at finite temperature has been an active topic of investigation,
that led to a variety of methods for its calculation.\citep{ichiye_91,amadei_93,ramirez01,wheeler_03,schmitz_04,martinez_06,turney_09,thomas_10,koukaras_15,ravichandran_18}
In principle, molecular dynamics (MD) simulations, that generate a
time trajectory in the phase space of positions and momenta, may seem
to be more appropriate than Monte Carlo (MC) methods for the analysis
of vibrational modes. MC simulations are based on a random walk exploration
of the configuration space of positions. This difference may be the
reason why the most employed methods to study vibrational modes in
equilibrium at finite temperatures are based on the analysis of Fourier
transformed velocity time-correlation functions in MD simulations.\citep{martinez_06,thomas_10,koukaras_15}
Other alternative methods seek to describe anharmonic shifts in vibrational
frequencies in a lattice dynamics framework by considering third and
fourth order force constants,\citep{turney_09} and renormalization.\citep{ravichandran_18}

Nevertheless, there are also efficient approaches to study collective
vibrations that work on the configuration space of positions, being
thus applicable to either classical MD or MC simulations. They are
based on the study of the covariance matrix of the atomic displacements.
The first applications of these methods were the analysis of collective
motions,\citep{ichiye_91} and the so-called essential dynamics in
proteins.\citep{amadei_93} The principal mode analysis (PMA), based
also on the study of spatial correlations by their covariance matrix,
was applied to the study of molecular vibrations in condensed phases.\citep{wheeler_03,wheeler_03b,schmitz_04}
Essentially the same method with focus in the calculation of phonon
dispersion relations in periodic solids was formulated in Dove's book
on lattice dynamics. \citep{dove} A common characteristic of all
these approaches is that they are applied within the framework of
MC or MD equilibrium simulations in the classical limit.

The so-called harmonic linear response (HLR) analysis of vibrational
problems is also based on the calculation of spatial correlations
in the configuration space of positions, but in contrast to all the
previous methods it was formulated within the framework of equilibrium
quantum path integral (PI) simulations.\citep{ramirez01} The PI formulation
of statistical mechanics shows how a quantum system (here, molecule
or periodic solid) can be mapped onto a classical model of interacting
``ring polymers''.\citep{feynman_65,feynman_86,ceperley95} An interesting
concept in this mapping is the centroid density, i.e. the spatial
probability density of the centroid coordinate. This coordinate is
the center of mass of the ring polymer that represents a given quantum
particle. In the classical limit, the ring polymer of the path integral
formalism collapses spatially into its centroid, so that the centroid
density becomes identical to the classical spatial probability density
of the particle. Thus, in this limit, the HLR approach becomes closely
related to the already mentioned methods of studying vibrational modes
by means of the covariance matrix of atomic displacements.\citep{ichiye_91,ramirez01,amadei_93,dove,wheeler_03,wheeler_03b} 

The HLR approach was derived by considering the statistical mechanics
of linear response and the fluctuation-dissipation theorem.\citep{kubo_91}
The latter formulates that the static response of a system in equilibrium
to an external disturbance (here, an external force) is a function
of the spontaneous fluctuation in the conjugate variable (here, the
centroid displacement) in absence of the force. The HLR analysis of
atomic fluctuations using the centroid variable displays a broad generality,
in the sense that it can be applied to both MD and MC simulations,
as well as to quantum and classical ones. Quantum vibrational energies
of both molecules and crystals have been investigated by this method
so far.\citep{lopez_03,ramirez05,ramirez_06,herrero_06,herrero09,herrero10}
However, the previous HLR treatment of periodic systems did not exploit
the translational symmetry of the atoms within the simulation cell,\citep{ramirez05}
a limitation that will be overcome in this paper.

The main purpose of the present work is then to investigate the capability
of the HLR method in the study of the phonon dispersion of two-dimensional
(2D) solids. Since the experimental characterization of graphene as
a one-atom thick solid membrane,\citep{novoselov04,novoselov05} a
huge amount of experimental and theoretical work has been devoted
to 2D materials.\citep{amorim14,roldan17} In particular, several
MC simulations focused on the study of surface ripples on graphene
by the analysis of the Fourier transform of the correlation function
of the out-of-plane 
displacements.\citep{fasolino07,liu09,Los09,zakharchenko10,roldan11,hasik18}
The interest of the HLR study of 2D solids, such as graphene, is to
highlight several facts that, to a large extent, have been overlooked
in previous simulations of this material: $i)$ the close relationship
between the Fourier transform of correlation functions of out-of-plane
coordinates and the phonon dispersion of the material; $ii)$ the
complete phonon dispersion of the 2D layer can be derived by extending
the analysis of out-of-plane fluctuations to the other in-plane coordinates.
A necessary step for this goal is to formulate the HLR approach using
symmetry adapted coordinates in solids with 2D periodicity. 

The present paper is organized as follows. Sec. \ref{sec:HLR-Method}
focuses on the formulation of the HLR method to derive phonon dispersion
relations of 2D solids. It is divided into three Subsections: A) an
introduction of the HLR approach using a single particle moving in
a 1D potential; B) the generalization of the method to the study of
2D layers in thermal equilibrium by either quantum PI or classical
simulations; C) the treatment of reciprocal space. Several applications
to illustrate the capability of the HLR method to calculate phonon
dispersion relations of 2D solids are presented. 
Sec. \ref{sec:HLR-Applications:graphene}
is devoted to graphene, while Sec. \ref{sec:Applications:-graphene-bilayer}
deals with a graphene bilayer and graphane. Our main interest is to
check the internal consistency of the HLR approach as a tool to analyze
spatial atomic fluctuations, derived by either classical or quantum
path integral simulations. Therefore, whenever possible, we will compare
HLR predictions of physical quantities that may be derived by other
independent methods. The applications include the phonon dispersion
of a graphene monolayer, a graphene bilayer and a graphane layer.
In the case of graphene, the temperature dependence of several observables
related to the anharmonicity of the layer have been also analyzed,
namely, the atomic kinetic energy, the frequency of the $E_{g}$ optical
modes, and the elastic moduli. The paper closes with a summary.

\section{The HLR Method\label{sec:HLR-Method}}

In this section, the HLR approach is introduced first for a single
particle in an anharmonic potential, and then applied to the study
of 2D solids by exploiting their translational symmetry.

\subsection{Linear response to an external constant force }

Spatial probability densities of particles in thermal equilibrium
carry information about their linear response to applied external
forces. For the sake of clarity, a quantum particle having bound states
in a one-dimensional potential, $U(x)$, is considered in equilibrium
at temperature $T$. The linear change of its average 
position $\left\langle x(f)\right\rangle $
under the application of an external constant force $f$ is given
by\citep{ramirez01}
\begin{equation}
\triangle x = \left\langle x(f)\right\rangle -\left\langle x(0)\right\rangle =
	    \beta\delta X^{2}f+O(f^{2}) \:,  \label{eq:Hooke}
\end{equation}
where the brackets $\left\langle \:\right\rangle $ indicate a thermal
average, $\beta=1/k_{B}T$ is the inverse temperature and $\delta X^{2}$
is the dispersion of the centroid density of 
the particle.\citep{feynman_86,cao_90,voth_91}
This quantity is readily obtained from equilibrium path integral simulations
of the unperturbed ($f\equiv0$) system. In the classical limit for
an arbitrary potential $U$, the dispersion $\delta X^{2}$ becomes
identical to the dispersion of the spatial probability density of
the particle in the potential $U$, i.e., 
\begin{equation}
\delta X^{2}\underset{cla}{\equiv}\delta x^{2} = 
       \left\langle x^2 \right\rangle -\left\langle x\right\rangle ^2  \:.
\end{equation}
This relation, $\delta X^{2}=\delta x^{2}$, is also valid for a quantum
particle in a harmonic potential, $U_{H}$, but anharmonicities in
$U$ make that, for quantum systems, $\delta X^{2}$ and $\delta x^{2}$
become different.\citep{ramirez_99b}

The shift $\triangle x$ of the average position of the particle with
respect to an external force $f$ in Eq. (\ref{eq:Hooke}) resembles
the Hooke's law of a mechanical system ($\triangle x=f/k_{eff}$),
with the proportionality constant playing the role of the inverse
of an effective force constant $k_{eff}$ of the particle vibrating
in the potential $U$. Thus the angular frequency $\omega$ of the
first vibrational excitation of the particle in an anharmonic potential
may be approximated as a function of the centroid density as
\begin{equation}
m\omega^{2}=k_{eff}=\frac{1}{\beta\delta X^{2}}\:,\label{eq:HLR-1D}
\end{equation}
with $m$ being the particle mass. This expression is the harmonic
linear response (HLR) approximation to vibrational excitation 
energies.\citep{ramirez01}
This approach is different from a standard harmonic approximation
(HA). The HA is temperature independent and predicts a vibrational
frequency, $\omega_{H}$, that assuming the position $x=0$ as the
potential energy minimum, is given as
\begin{equation}
m\omega_{H}^{2}=\left(\frac{\partial^{2}U}{\partial x^{2}}\right)_{x=0}\:.
\end{equation}
The HLR approach is sensitive to anharmonic effects that are neglected
in a HA.\citep{ramirez01} Note that the coefficient $k_{eff}$ in
Eq. (\ref{eq:HLR-1D}) defines a linear response of the particle in
the potential $U$. Thus, in words Eq. (\ref{eq:HLR-1D}) implies
that, a fictitious particle moving in the (temperature dependent)
harmonic potential $U_{H}=k_{eff}x^{2}/2$ would display identical
linear response in thermal equilibrium as the true particle in the
anharmonic potential $U$. 

The extension of the HLR approach to a many-body vibrational system
is straightforward.\citep{ramirez01} Nevertheless, a previous treatment
of periodic systems did not exploit the translational symmetry within
the simulation supercell.\citep{ramirez05} Then, the whole set of
vibrational frequencies, $\left\{ \omega_{j}\right\} $, were derived
as $\mathbf{k}=0$ states of the (small) folded Brillouin zone (BZ)
of the employed supercell. This limitation is overcome in the next
Subsections for the case of solids with 2D periodicity, allowing to
assign to each of the computed frequencies, $\omega_{j}$, its corresponding
$\mathbf{k}$-vector within the (larger) unfolded BZ of the primitive
lattice. 

\subsection{Treatment of 2D solids}

For the sake of generality, a $NPT$ simulation with a flexible 2D
supercell defined with translation vectors $(\mathbf{a}_{1,}\mathbf{a}_{2})$
is considered. Both $NVT$ or $NPT$ simulations with isotropic volume
changes are particular cases whose treatment is easily derived from
the general one. We assume that global translations and global rotations
of the simulation cell are excluded from the stored trajectory, i.e.,
all atomic displacements in the simulation will correspond to vibrational
degrees of freedom. The absence of global translations is easily assured
by maintaining the center of mass of the simulation cell fixed along
the simulation run. Global rotations are not possible in $NVT$ or
isotropic $NPT$ simulations, as they are incompatible with the application
of periodic boundary conditions. In flexible $NPT$ simulations one
should check that the algorithm employed to sample cell fluctuations
does not allow for global cell rotations.\citep{ma96}

The output of any $NPT$ simulation is a trajectory showing the evolution
of the simulation supercell and the $N$ atomic positions along the
simulation. The equilibrium simulation cell is defined by a $2\times2$
matrix $\mathbf{\left\langle \mathbf{\Gamma}\right\rangle },$ whose
columns are the Cartesian coordinates of the cell axes,
\begin{equation}
\mathbf{a}_{i}=\mathbf{\left\langle \mathbf{\Gamma}_{\mathit{1i}}\right\rangle }
	\mathbf{\hat{e}}_{1}+\mathbf{\left\langle \mathbf{\Gamma}_{\mathit{2i}}
	\right\rangle }\mathbf{\hat{e}}_{2}\;,(i=1,2)\:,
\end{equation}
where $\left\{ \mathbf{\hat{e}}_{1},\mathbf{\hat{e}}_{2}\right\} $
represent unit vectors along the Cartesian axes and the brackets 
$\left\langle \:\right\rangle $
indicate an ensemble average. For a given observable $x$, 
$\left\langle x\right\rangle $
is estimated by an arithmetic mean over the generated trajectory,
\begin{equation}
\left\langle x\right\rangle =S^{-1}\sum_{s=1}^{S}x_{s}\:,
\end{equation}
where $S$ is the number of steps in the simulation run.

In a quantum PIMD or PIMC simulation, we assume that the set of atomic
centroid coordinates, $(\mathbf{r}_{i},z_{i})$ ($i=1,\ldots,N)$,
are stored along the simulation run, while in the case of a classical
MD or MC simulation, the coordinates $(\mathbf{r}_{i},z_{i})$ would
correspond to the atomic positions. $\mathbf{r}_{i}$ is a 2D vector
in the plane of the supercell, $(\mathbf{a}_{1},\mathbf{a}_{2})$,
while $z_{i}$ is the Cartesian coordinate in the perpendicular out-of-plane
direction. 

It is convenient to define 2D atomic fractional coordinates, $\mathbf{c}_{i},$
using the cell axes as basis. At a given simulation step with a cell
defined by $\mathbf{\Gamma}$, the fractional coordinates of the $i'$th
atom are
\begin{equation}
\mathbf{c}_{i}=\mathbf{\Gamma}^{-1}\mathbf{r}_{i}\:.
\end{equation}
$\mathbf{\Gamma}$ fluctuates along a $NPT$ simulation, but will
remain constant in a $NVT$ simulation. The average in-plane structure
is calculated as
\begin{equation}
\mathbf{r}_{eq,i}=\left\langle \mathbf{\Gamma}\right\rangle 
	 \left\langle \mathbf{c}_{i}\right\rangle \:.
\end{equation}

\begin{figure}
\vspace{0.1cm}
\hspace{-0.5cm}
\includegraphics[width=5.0cm]{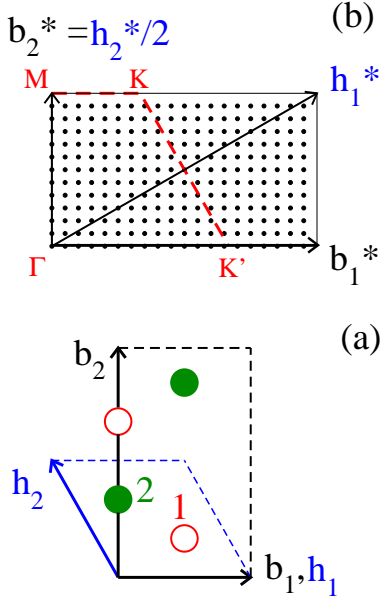}
\vspace{0.4cm}
\caption{(a) The axes $(\mathbf{b}_{1},\mathbf{b}_{2})$ display a centered
rectangular cell of a graphene layer containing four atoms. The primitive
hexagonal axes are $(\mathbf{h}_{1,}\mathbf{h}_{2})$. There are 2
types of basis atoms, represented by closed and open circles, respectively.
(b) The discrete $\mathbf{k}$-grid {[}see Eq. (\ref{eq:k-grid})
{]} used in the FT of Eq. (\ref{eq:FT}) corresponding to a supercell
$(\mathbf{a}_{1},\mathbf{a}_{2})=(20\mathbf{b}_{1},12\mathbf{b}_{2})$
is displayed by closed circles. $(\mathbf{b}_{1}^{*},\mathbf{b}_{2,}^{*})$
are the reciprocal basis vectors of the centered lattice, while $(\mathbf{h}_{1,}^{*},\mathbf{h}_{2,}^{*})$
are those of the primitive lattice. The reciprocal basis vectors of
the simulation supercell are $(\mathbf{a}_{1}^{*},\mathbf{a}_{2}^{*})=(\mathbf{b}_{1}^{*}/20,\mathbf{b}_{2}^{*}/12).$
The points $\varGamma,M,K$, and $K'$ are symmetry points at the
boundary of the hexagonal BZ. The $\mathbf{k}$-points to the right
of the $K-K'$-line lie outside the first BZ. A translation by the
reciprocal lattice vector 
$-\mathbf{h}_{1}^{*}=-(\mathbf{b}_{1}^{*}+\mathbf{b}_{2}^{*})$
moves these points within the first BZ.}
\label{fig:cell}
\end{figure}

The dynamic information in the vibrational problem is derived here
from the spatial atomic displacements $(\mathbf{R}_{i},Z_{i})$ along
the simulation run. These displacement coordinates are spatial deviations
with respect to the equilibrium average coordinates
\begin{equation}
\mathbf{R}_{i}=\left\langle \mathit{\mathbf{\Gamma}}\right\rangle 
    \left(\mathbf{c}_{i}-\left\langle \mathbf{c}_{i}\right\rangle \right)\:,
\end{equation}
\begin{equation}
Z_{i}=z_{i}-\left\langle z_{i}\right\rangle \:.
\end{equation}
 The next step for the derivation of the vibrational dispersion relations,
$\omega_{j}(\mathbf{k})$, is to exploit the translational symmetry
of the average 2D crystal structure. To this aim, the set of $N$
atoms in the simulation supercell, $\left\{ \mathbf{r}_{i}\right\} ,$
is divided into disjoint basis subsets, 
$\left\{ \mathbf{r}_{j}\right\} _{\alpha}$,
with $\alpha=1,\ldots n$, where $n$ is the number of basis atoms,
i.e., the number of atoms within the crystallographic primitive cell.
The number of atoms is identical for all subsets. It is equal to the
number of primitive cells, $P_{c}$, within the simulation supercell,
i.e., $P_{c}=N/n$. E.g., for the 2D hexagonal structure of a graphene
layer, the number of basis atoms is $n=2$. (see Fig. \ref{fig:cell}a).
Thus, in a simulation supercell with $N$ atoms, the carbon atoms
will be classified into two disjoint subsets, $\left\{ \mathbf{r}_{j}\right\} _{1}$
and $\left\{ \mathbf{r}_{j}\right\} _{2}$, each one with $N/2$ elements.
Note that, in the average crystal structure, $\left\{ \mathbf{r}_{eq,i}\right\} $,
the $P_{c}$ atoms of a given subset $\left\{ \mathbf{r}_{eq,j}\right\} _{\alpha}$
are all symmetry equivalent by the application of an appropriate primitive
translation of the crystal lattice. 

It is now convenient to relabel the displacement coordinates of the
$i'$th atom to indicate the type of basis atom. Then the subindex
$i$ ($i=1,\ldots N)$ will be replaced by a double subindex $\alpha j$,
\begin{equation}
(\mathbf{R}_{i},Z_{i})\longrightarrow(\mathbf{R}_{\alpha j},Z_{\alpha j})
	\equiv(X_{\alpha j},Y_{\alpha j,}Z_{\alpha j})\:,
\end{equation}
 where $\alpha$ ($\alpha=1,\ldots n)$ indicates that the $i'$th
atom belongs to the $\alpha'$th basis subset, and $j$ ($j=1,\ldots,P_{c})$
is a running index for the $P_{c}$ symmetry equivalent atoms of the
subset.

A symmetry adapted Bloch function, $\overline{X}_{\alpha}(\mathbf{k})$,
is a collective variable defined as a linear combination of the mass-weighted
displacement coordinates of the atoms within the $\alpha'$th 
subset,\citep{kittel,dove} 
\begin{equation}
\overline{X}_{\alpha}(\mathbf{k})=\sqrt{\frac{m_{\alpha}}{P_{c}}}
    \sum_{j=1}^{P_{c}}X_{\alpha j}\exp\left(i\mathbf{k}\mathbf{r}_{eq,\alpha j}
	\right)\:.\label{eq:FT}
\end{equation}
$m_{\alpha}$ is the atomic mass of an atom in the $\alpha'$th subset.
The number of Bloch functions is $3n$, i.e., 
$\left[\overline{X}_{1}(\mathbf{k}),\overline{Y}_{1}(\mathbf{k}),\overline{Z}_{1}(\mathbf{k}),\ldots,\overline{X}_{n}(\mathbf{k}),\overline{Y}_{n}(\mathbf{k}),\overline{Z}_{n}(\mathbf{k})\right]$,
that corresponds to the number of vibrational bands in the 2D solid.
Note that the phase factor in the discrete Fourier transform (FT)
is defined by the average crystal structure, $\left\{ \mathbf{r}_{eq,\alpha j}\right\} $.
This phase factor does not change when the coordinate $X_{\alpha j}$
is replaced by any other coordinate, $Y_{\alpha j}$ or $Z_{\alpha j}$,
to derive the corresponding FT. The next (and nearly last) step is
to calculate the hermitean susceptibility tensor, $\chi(\boldsymbol{\mathrm{k}})$.\citep{ramirez01}
Its $3n\times3n$ elements are defined as ensemble averages of products
of the form $\overline{A}_{\alpha}(\mathbf{k})\overline{B}_{\gamma}^{*}(\mathbf{k})$,
with $\alpha$ and $\gamma$ as indices labeling basis atoms, and
$A$ and $B$ as letters labeling any of the $(X,Y,Z)$ displacement
coordinates. These products are the covariance of symmetry adapted
functions of displacement coordinates. They provide a quantitative
measure of the correlations in the atomic vibrations of the solid.
A diagonal element of the tensor $\chi(\boldsymbol{\mathrm{k}})$
is derived as 
\begin{equation}
\left\langle \overline{X}_{\alpha}(\mathbf{k})\overline{X}_{\alpha}^{*}(\mathbf{k})
   \right\rangle =S^{-1}\sum_{s=1}^{S}|\overline{X}_{\alpha}(\mathbf{k})|_{s}^{2}\:,
\label{eq:dia_dispersion}
\end{equation}
where $s$ is a running index for the simulation steps, and the vertical
bars denote the modulus of the complex number. A non-diagonal element
of $\chi(\boldsymbol{\mathrm{k}})$ is calculated as
\begin{equation}
\left\langle \overline{X}_{\alpha}(\mathbf{k})\overline{Y}_{\beta}^{*}(\mathbf{k})\right\rangle =S^{-1}\sum_{s=1}^{S}\left(\overline{X}_{\alpha}(\mathbf{k})\overline{Y}_{\beta}^{*}(\mathbf{k})\right)_{s}\:.\label{eq:nondia_dispersion}
\end{equation}
The last step is the diagonalization of the hermitean tensor $\chi(\boldsymbol{\mathrm{k}})$
that yields $3n$ real eigenvalues $\triangle_{j}(\mathbf{k})$. The
eigenvectors of the tensor $\chi(\mathbf{k})$ are the Cartesian displacement
vectors of the vibrational modes of the crystal, while the eigenvectors
$\triangle_{j}(\mathbf{k})$ are a measure of the spatial dispersion
of each mode. The one-body HLR result in Eq. (\ref{eq:HLR-1D}), that
relates the spatial dispersion $\delta X^{2}$ with the angular frequency
$\omega$, can be generalized to the many-body problem to obtain the
phonon dispersion bands $\omega_{j}(\mathbf{k})$ of the 2D solid
as\citep{ramirez01}
\begin{equation}
\omega_{j}^{2}(\mathbf{k})=\frac{1}{\beta\varDelta_{j}(\mathbf{k})}\:.
\end{equation}

\subsection{Treatment of reciprocal space\label{subsec:Treatment-of-k-space} }

The reciprocal lattice vectors $(\mathbf{a}_{1}^{*},\mathbf{a}_{2}^{*})$
of the equilibrium simulation supercell are the columns of the matrix
$\mathbf{S}$ defined as
\begin{equation}
\mathbf{S}=2\pi\left(\mathbf{\left\langle \Gamma\right\rangle }^{-1}\right){}^{T}\:,
\end{equation}
so that $\mathbf{a}_{i}\mathbf{a}_{j}^{*}=2\pi\delta_{ij}$, with
$\delta_{ij}$ as the Kronecker delta. The discrete FT in Eq. (\ref{eq:FT})
is defined for $\mathbf{k}$-vectors that are commensurate with the
simulation supercell, i.e., 
\begin{equation}
\mathbf{k}=k_{1}\mathbf{a}_{1}^{*}+k_{2}\mathbf{a}_{2}^{*}=(k_{1,}k_{2})\:.
\label{eq:k-grid}
\end{equation}
The rhs gives the components of $\mathbf{k}$ using $(\mathbf{a}_{1}^{*},\mathbf{a}_{2}^{*})$
as basis vectors, with $k_{1}$ and $k_{2}$ being integers. If the
simulation cell is defined as a $M_{1}\times M_{2}$ supercell of
a crystallographic unit cell (either primitive or non-primitive),
with $M_{1}$ and $M_{2}$ as positive integers, then $k_{1}$ and
$k_{2}$ will vary between $0\leq k_{1}\leq M_{1}-1,$ and $0\leq k_{2}\leq M_{2}-1$. 

For some applications it is convenient to convert the $\mathbf{k}$-grid
of Eq. (\ref{eq:k-grid}) into a symmetry equivalent one, that is
located within the first BZ of the lattice. In the case that the crystallographic
unit cell is a primitive one, the set of $\mathbf{k}$-vectors, $\mathbf{g}_{0}=(0,0)$,
$\mathbf{g}_{1}=(M_{1,}0)$, $\mathbf{g}_{2}=(0,M_{2})$, and $\mathbf{g}_{3}=(M_{1},M_{2})$,
are reciprocal lattice vectors of the primitive lattice. Then, the
four vectors $\{\mathbf{k}-\mathbf{g}_{0}$, $\mathbf{k}-\mathbf{g}_{1}$,
$\mathbf{k}-\mathbf{g}_{2}$, $\mathbf{k}-\mathbf{g}_{3}\}$ correspond
to symmetry equivalent points in reciprocal space. Only one point
of this set will belong to the first BZ. A sufficient condition  for
$\mathbf{k}$ to be located within the BZ is to choose the point whose
vector modulus is the smallest, i.e.,
\begin{equation}
\mathbf{k}\equiv\mathrm{min\;mod}\left\{ \mathbf{k}-\mathbf{g}_{0},\mathbf{k}-\mathbf{g}_{1},\mathbf{k}-\mathbf{g}_{2},\mathbf{k}-\mathbf{g}_{3}\right\} \:.\label{eq:k_in_BZ}
\end{equation}
If the employed crystallographic unit cell is a non-primitive centered
lattice, the vector from the set in Eq. (\ref{eq:k_in_BZ}) that is
located within the first BZ is determined just by making a sketch
in reciprocal space of the relative position of the grid of $\mathbf{k}$-vectors
in Eq. (\ref{eq:k-grid}) and the BZ of the primitive lattice. See
Fig. \ref{fig:cell} for a worked example of a graphene layer using
a non-primitive centered rectangular cell.

\section{applications: graphene \label{sec:HLR-Applications:graphene}}

The selected applications are intended as a test to illustrate the
capability of the HLR approach in the calculation of phonon dispersion
relations of 2D solids. Our main interest is to show the internal
consistency of the HLR approach by comparing its predictions with
other independent methods, and using a graphene monolayer as main
example. Our vibrational analysis requires only spatial positions
and it is equally applicable to classical and quantum simulations
performed by either MD or MC methods.\textcolor{red}{{} }The fact that
the HLR analysis can be applied indistinctly to either classical or
quantum path integral simulations is an important feature of the method.
To illustrate this capability, the selected applications of the HLR
approach for graphene include both classical as well as path integral
simulations.

\subsection{Harmonic phonon dispersion in graphene \label{subsec:Graphene}}

A simulation of the harmonic limit of an anharmonic potential must
be performed in the classical low temperature limit. In the quantum
case, anharmonic effects related to zero-point vibrations appear even
in the $T\rightarrow0$ limit.\citep{herrero16} Under low temperature
conditions, the agreement of the HLR phonon dispersion, $\omega(\mathbf{k}),$
and the harmonic result, $\omega_{H}(\mathbf{k})$, will serve as
test of the HLR method. $\omega_{H}(\mathbf{k})$ is calculated, with
independence from the simulation, by diagonalization of the vibrational
dynamical matrix.\citep{kresse_1995} 

\subsubsection{Computational conditions}

The empirical long-range carbon bond order potential (LCBOPII) was
employed in this calculation.\citep{los05} In line with previous
simulations a slight modification of the original torsion parameters
was made to increase the bending constant of a flat layer in the $T\rightarrow0$
limit from $\kappa=$0.82 eV to a more realistic value of $\kappa=$1.48
eV,\citep{ramirez16,tisi_17} closer to experimental data and ab-initio
calculations.\citep{lambin14} A temperature of 1 K and vanishing
external stress ($P$=0) were chosen for the graphene MD simulation
in the isotropic $NPT$ ensemble. The classical MD simulation was
performed with a time step of 1 fs. The equilibration run consisted
on $2\times10^{6}$ MD steps (MDS) and a trajectory with 48000 layer
configurations was stored at equidistant steps from a run with $2.4\times10^{7}$
MDS. The stored trajectory was used to calculate the covariance of
the atomic probability distributions by Eqs. (\ref{eq:dia_dispersion})
and (\ref{eq:nondia_dispersion}). From a non-primitive centered rectangular
graphene cell with 4 carbon atoms (see Fig. \ref{fig:cell}a), a $20\times12$
supercell with $N=960$ atoms was defined for the graphene simulations.
This rectangular supercell displays similar lengths in the $(x,y)$-directions.
The dimension of the tensor $\chi(\mathbf{k})$ is $6\times6$ for
graphene. The shortest distance between points of the $\mathbf{k}$-grid
of Eq. (\ref{eq:k-grid}) is 
$\left|\mathbf{a}_{1}^{*}\right|\simeq\left|\mathbf{a}_{2}^{*}\right| 
  = 0.12$ \AA$^{-1}$
(see Fig. \ref{fig:cell}b).

\subsubsection{Phonon dispersion}

\begin{figure}
\vspace{0.1cm}
\includegraphics[width=7.0cm]{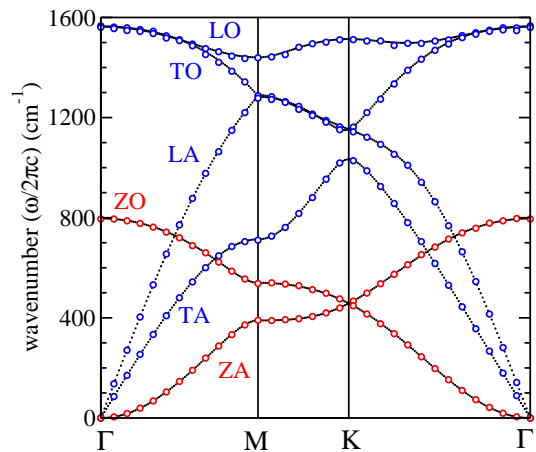}
\vspace{0.6cm}
\caption{The harmonic phonon dispersion of graphene, $\omega_{H}(\mathbf{k}$),
as derived by diagonalization of the dynamical matrix of the LCBOPII
potential, is shown by dotted lines that look like solid lines when
the slope is small. The six phonon bands are classified by their labels.
The open circles are the HLR result as derived from the spatial atomic
fluctuations obtained by a classical $NPT$ simulation with the same
potential model. The MD simulation was performed close to the harmonic
$T\rightarrow0$ limit, at 1 K and $P=0$, on a $20\times12$ supercell
of a centered rectangular cell of graphene (see Fig. \ref{fig:cell}a).
The $6\times6$ susceptibility tensor $\chi(\mathbf{k})$ displays
a block structure with separation of $(X,Y$) (blue circles) and $Z$
(red circles) bands. }
\label{fig:w_k_graphene}
\end{figure}

The LCBOPII phonon dispersion of graphene, $\omega_{H}(\mathbf{k})$,
derived from the diagonalization of the dynamic matrix, is presented
in Fig. \ref{fig:w_k_graphene}. The hexagonal cell parameter amounts
to 2.4593 \AA.  The result corresponds to a dense set
of points along the boundaries of the irreducible BZ (IBZ) of the
hexagonal lattice. The dispersion relation of graphene comprises three
acoustic (A) and three optical (O) bands, which are either in-plane
longitudinal (L), in-plane transverse (T) or out-of-plane (Z). The
acoustic ZA mode displays a $k^{2}$-dependence near $\mathit{\Gamma}$,
in contrast with the linear dispersion of the TA and LA modes, which
is typical for acoustic modes in 3D solids.\citep{zimmermann_08}
The $\omega_{H}(\mathbf{k})$ bands in Fig. \ref{fig:w_k_graphene}
(lines) are in agreement to previous calculations using the same LCBOPII
model.\citep{locht_12}

The HLR result derived from the diagonalization of the $\chi(\mathbf{k})$
tensor are displayed as open circles in Fig. \ref{fig:w_k_graphene}.
The displayed $\mathbf{k}$-points correspond to those points in Eq.
(\ref{eq:k-grid}) that lie at the boundary of the IBZ. The agreement
with the $\omega_{H}(\mathbf{k})$ dispersion is excellent for the
six vibrational bands. The calculation of the tensor $\chi(\mathbf{k})$
includes the constraint provided by translational symmetry, but not
from point symmetry elements of the crystal. Then, the symmetry of
the vibrational modes will display errors associated to the finite
sampling along the simulation. E.g., this error is about 0.3 \% ($\sim$6
cm$^{-1}$) for the degenerate LO and TO modes at $\varGamma$. One
could impose additional symmetry constraints (rotational axes and
reflection planes) to the tensor $\chi(\mathbf{k})$ to ensure that
the vibrational modes display the correct symmetry and smaller statistical
errors. The reducible representation of the displacement vectors $(X_{\alpha j},Y_{\alpha j},Z_{\alpha j})$
of atoms in a solid is the same as that of atomic $p$-orbitals. Thus,
the method to symmetrize $\chi(\mathbf{k})$ would be identical to
that already used for the density matrices associated to $p$-orbitals
in electronic band structure calculations based on crystal orbital
methods.\citep{pisani_80,ramirez_88} 

In the next Subsections, anharmonic shifts predicted by the HLR approach,
caused by quantum zero-point motion and/or by an increase in temperature,
are analyzed showing that they do lead to a consistent description
of the collective vibrations.

\subsection{Atomic kinetic energy in graphene}

The vibrational kinetic energy of atoms in solids or molecules is
a physical quantity that is affected by both quantum and anharmonic
effects.\citep{ramirez_11} The kinetic energy of the carbon atoms
in graphene can be obtained by quantum PI simulations using the virial
estimator. \citep{herman82} This non-perturbative approach allows
for an, in principle, ``exact'' treatment of both quantum and anharmonic
effects as a function of temperature. This quantum observable is used
to test the predictions of the HLR approach. In the classical limit,
the vibrational kinetic energy amounts to $k_{B}T/2$ per degree of
freedom (equipartition theorem), a value that does not depend on the
anharmonicity of the interatomic potential.

\subsubsection{Computational conditions}

Quantum PIMD simulations of graphene with the LCBOPII model in the
isotropic $NPT$ ensemble were performed at discrete temperatures
between 25 and 750 K for an unstressed layer ($P=0$). The simulation
cell with $N=960$ atoms is the same as that one described in Subsec.
\ref{subsec:Graphene}. The Trotter number, $N_{Tr}$, that characterizes
the applied discretization of the path integral was set as a temperature
dependent value by the relation $N_{Tr}T=6000$ K. At 25 K, the lowest
studied temperature, $N_{Tr}=240$, while at 750 K, the highest studied
temperature, $N_{Tr}=8$. The time step was set as 0.5 fs, equilibration
and simulations runs comprise $5\times10^{5}$ MDS and $8\times10^{6}$
MDS, respectively. At each temperature, the susceptibility tensor
$\chi(\mathbf{k})$ was calculated from a set of $1.6\times10^{4}$
crystal configurations stored at equidistant steps along the run.
Further simulation conditions are identical to those already described
in our previous PIMD graphene simulations.\citep{herrero16,herrero18}

\subsubsection{Kinetic energy in the HLR approximation}

\begin{figure}
\vspace{0.1cm}
\includegraphics[width=7.0cm]{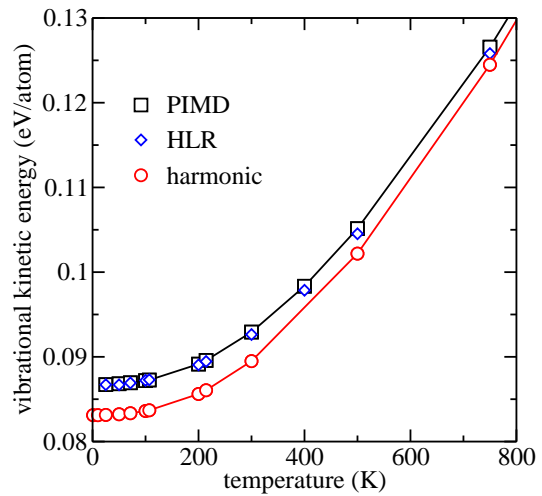}
\vspace{0.6cm}
\caption{Vibrational kinetic energy of the carbon atoms in graphene as a function
of temperature. The ``exact'' quantum PIMD result derived from the
virial estimator (open squares) is compared to the approximation of
Eq. (\ref{eq:K_LR}) using the set of vibrational frequencies, $\omega(\mathbf{k})$,
obtained either by the HLR approach (open diamonds) or in the harmonic
approximation (open circles). The lines are guides to the eye.}
\label{fig: kin_LR_temp}
\end{figure}

The kinetic energy per atom, as derived from the PIMD simulation,
is presented as a function of temperature in Fig. \ref{fig: kin_LR_temp}
as open squares. We observe that at low temperature the kinetic energy
converges toward a constant zero-point value, while as temperature
increases the thermal excitation of vibrational modes produces a gradual
increase of the kinetic energy that tends to linearity in the high
temperature limit.

The vibrational kinetic energy per atom can be estimated from the
HLR frequencies, $\omega_{j}(\mathbf{k})$, derived by diagonalization
of the susceptibility tensor, $\chi(\mathbf{k})$, as 
\begin{equation}
E_{K}=\frac{1}{nN_{k}}\sum_{i=1}^{N_{k}}\sum_{j=1}^{3n}\frac{\hbar\omega_{j}(\mathbf{k}_{i})}{4}\coth\left(\frac{\beta\hbar\omega_{j}(\mathbf{k}_{i})}{2}\right)\:,\label{eq:K_LR}
\end{equation}
where the number of basis atoms is $n=2,$ and the number of $\mathbf{k}$-points
is $N_{k}=$240 (as corresponds to the $20\times12$ simulation supercell,
see Subsec. \ref{subsec:Treatment-of-k-space}). Each mode of frequency,
$\omega_{j}(\mathbf{k})$, is assumed to contribute to the kinetic
energy by a quantum harmonic expression. Therefore anharmonicity in
this model appears only in the value of the frequencies, $\omega_{j}(\mathbf{k})$,
which may display shifts with respect to the harmonic limit. The HLR
estimation of the kinetic energy is displayed in Fig. \ref{fig: kin_LR_temp}
as open diamonds. We observe an excellent agreement to the ``exact''
kinetic energy derived by the virial estimator at temperatures below
300 K. Anharmonic effects are present in the whole temperature range,
but it is expected that they become larger as temperature rises. This
might be the reason for the slight underestimation of the ``exact''
value of $E_{K}$ by the HLR approximation at the highest studied
temperatures around 750 K. Even at this relatively high temperature,
the kinetic energy in the classical limit ($3k_{B}T/2=0.097$ eV/atom)
is significantly lower ($\sim$24\%) than the value derived in the
quantum approach.

The contribution of anharmonicity to the kinetic energy can be evaluated
by calculation of the harmonic limit, $E_{KH}$, of the employed model
potential. $E_{KH}$ is derived by replacing the HLR frequencies,
$\omega_{j}(\mathbf{k})$, in Eq. (\ref{eq:K_LR}) by harmonic ones,
$\omega_{H,j}(\mathbf{k})$. The harmonic values, $\omega_{H,j}(\mathbf{k})$,
which do not depend on temperature, were calculated by diagonalization
of the dynamical matrix for a layer corresponding to the minimum potential
energy (see Fig. \ref{fig:w_k_graphene}). The harmonic kinetic energy,
$E_{KH},$ shown as open circles in Fig. \ref{fig: kin_LR_temp},
differs appreciably from the ``exact'' $E_{K}$ in the whole temperature
range. The anharmonic effect predicted by the LCBOPII model is an
increase in the vibrational kinetic energy. The increase amounts to
about $5\%$ at low temperatures, and is lower at higher temperatures
($\sim2\%$ at 750 K). This behavior at high temperature is surely
due to a compensation of errors, as anharmonic effects are expected
to increase with temperature. The improved results of the HLR approach
with respect to the harmonic limit reveal that the HLR frequencies
are really sensitive to the anharmonic effect in the vibrational modes.

\subsection{Temperature dependence of the optical phonons in 
graphene\label{subsec:Temperature-dependence-of}}

\subsubsection{LCBOPII model}

\begin{figure}
\vspace{0.1cm}
\includegraphics[width=6.0cm]{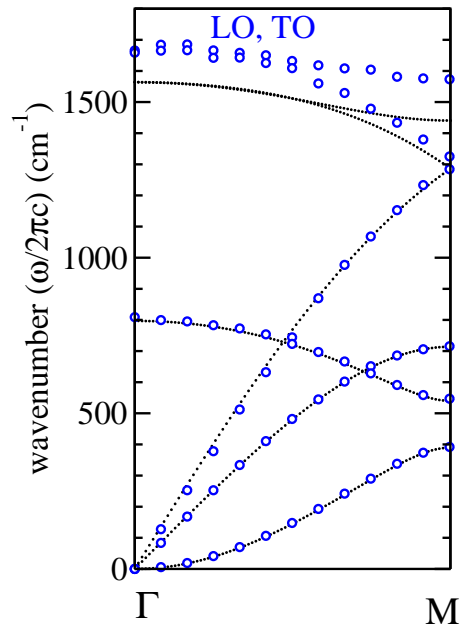}
\vspace{0.85cm}
\caption{The HLR result for the phonon dispersion of graphene derived from
a quantum PIMD simulation at 300 K is represented by open circles
along $\Gamma-\mathrm{M}$. The dotted lines show the harmonic limit
of the same model. The main anharmonic effect is a blue-shift of the
LO and TO bands. These results were derived with the LCBOPII model.}
\label{fig:w_gamma_M_300}
\end{figure}

The HLR phonon dispersion of graphene along the $\varGamma-M$ symmetry
direction, derived from the quantum PIMD simulations at 300 K, is
compared to the harmonic result in Fig. \ref{fig:w_gamma_M_300}.
We observe that the main anharmonic effect is a blue-shift of the
LO and TO bands that amounts to $\sim100$ cm$^{-1}$ ($\sim6\%$)
at the BZ center $\varGamma$. This anharmonic shift is the origin
for the increase in the vibrational kinetic energy of the C atoms
in the quantum PIMD simulation, in comparison to the quantum harmonic
limit, as shown in Fig. \ref{fig: kin_LR_temp}. Nevertheless, this
blue-shift seems to be an artifact of the LCBOP model, as it is in
disagreement to both first-principles calculations\citep{bonini_07}
and experimental results derived from Raman spectra.\citep{zhang_14} 

\begin{figure}
\vspace{0.1cm}
\includegraphics[width=7.0cm]{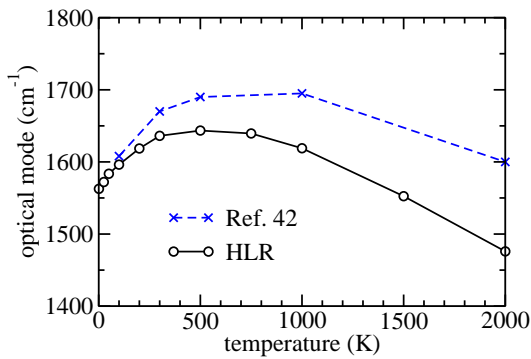}
\vspace{0.4cm}
\caption{Temperature dependence of the degenerate LO/TO phonon wavenumber of
graphene at $\varGamma$ as derived from classical simulations with
the LCBOPII model. The results of Ref. \onlinecite{tisi_17}, displayed
by crosses, were derived from the Fourier transformed velocity time
correlation functions. The results of the HLR analysis of classical
MD simulations are given as open circles. Lines are guides to the
eye. }
\label{fig:comp_hlr_velocity}
\end{figure}

An anharmonic blue-shift of the optical modes in graphene has been
reported in three recent studies of the temperature dependence of
its vibrational properties using either the LCBOP or LCBOPII potentials
in classical MD simulations.\citep{koukaras_15,anees_2015,tisi_17}
The results of the LO/TO wavenumber of graphene at $\varGamma$ calculated
in Ref. \onlinecite{tisi_17} from the Fourier transform of velocity
time correlation functions with the LCBOPII model is displayed as
a function of temperature in Fig. \ref{fig:comp_hlr_velocity} (crosses).
For comparison, the HLR result, as derived from classical simulations
with the same potential model, is displayed as circles. A similar
non-monotonic temperature dependence of the anisotropic shift of this
optical mode is predicted by both methods, although there appear quantitative
differences in the LO/TO wavenumbers, specially at high temperatures. 

\subsubsection{Tight-binding model}

We have re-analyzed the temperature dependence of the frequency of
the degenerate LO and TO optical phonons of graphene at $\varGamma$
using an efficient tight-binding (TB) Hamiltonian parameterized using
density-functional calculations,\citep{porezag_95} as an improved
alternative to the LCBOPII model. The symmetry of this Raman active
mode is $E_{2g}$ and corresponds to the G peak in the Raman 
spectra.\citep{ferrari_13}

\paragraph{Computational conditions:}

Quantum and classical simulations of graphene were performed in the
$NPT$ ensemble with isotropic volume fluctuations at $P=0$. The
electronic structure has been treated with an efficient tight-binding
(TB) Hamiltonian parameterized using density-functional 
calculations.\citep{porezag_95}
The temperature was varied between 50 and 1000 K in the quantum PIMD
simulations, and between 1 and 1000 K in the classical MD simulations.
The time step was taken between 0.25 and 1 fs. The lowest time step
of 0.25 fs was required for the quantum simulations at 50 K, where
the atomic spatial delocalizations related to the zero-point quantum
fluctuations are the largest. The simulation cell included 60 carbon
atoms with periodic boundary conditions and the electronic energy
was obtained using only the $\mathit{\Gamma}$-point for the calculation
of the crystal band structure. The Trotter number, $N_{Tr}$, was
set as a temperature dependent value by the relation $N_{Tr}T=6000$
K. Simulation runs comprise $8\times10^{6}$ MDS and the analysis
of the atomic real space fluctuations was performed on a subset of
$1.6\times10^{4}$ configurations stored equidistantly along the whole
trajectory.

\paragraph{HLR analysis of the LO/TO phonons:}

\begin{figure}
\vspace{0.1cm}
\includegraphics[width=7.0cm]{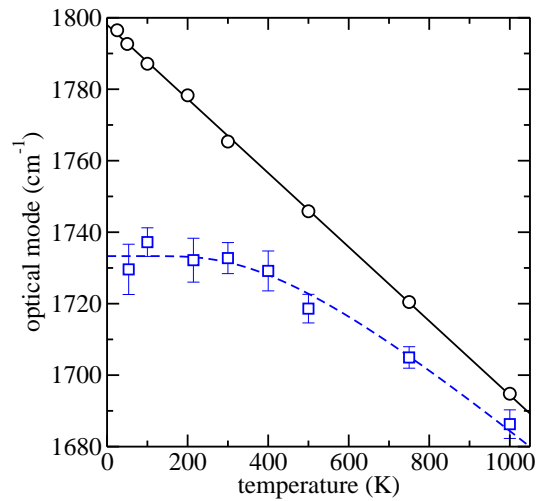}
\vspace{0.4cm}
\caption{Temperature dependence of the $E_{2g}$ degenerate LO/TO vibrational
states of graphene at $\varGamma$. This state corresponds to the
G peak in the Raman spectrum. The results were derived by the HLR
analysis of classical MD (open circles) and quantum PIMD (open squares)
trajectories using a TB Hamiltonian. The continuous line is a linear
fit of the classical result. The broken line is the least-squares
fit of the quantum result to Eq. (\ref{eq:fit_delta_w}).}
\label{fig:e2g_temp}
\end{figure}

The temperature dependence of the optical LO and TO vibrational modes
at $\varGamma$ as derived from the HLR method is displayed in Fig.
\ref{fig:e2g_temp}. In the classical limit the optical mode display
a linear temperature dependence. The classical $T\rightarrow0$ limit
corresponds to the harmonic expectation, $\omega_{H}=1798$ cm$^{-1}$
, for the TB model under the employed computational conditions. This
value overestimates by about 12\% the experimental value.\citep{zhang_14}
Vibrational frequencies determined from ab-initio calculations are
often scaled by empirical factors to compensate for systematic biases
that overestimate frequencies by about 10\% (e.g. in Hartree-Fock
6-31G(d) models).\citep{irikura_05} 

In the classical limit, the change of the optical mode frequency with
temperature is given by a slope of $\varphi_{T}=-0.1$ cm$^{-1}$/K.
The negative sign of the slope indicates that the anharmonicity of
the optical modes predicted by the TB method is indeed a red-shift
of the vibrational frequency, as opposed to the results derived previously
for the LCBOPII model. The quantum results for the temperature dependence
of this optical mode differ from the classical ones, specially below
room temperature. The intrinsic anharmonicity of the zero-point vibration
of the carbon atoms makes that the optical modes display a shift of
$\sim-64$ cm$^{-1}$ with respect to the classical expectation in
the low temperature limit. 

Temperature-dependent Raman scattering experiments have shown that
the temperature variation of the phonon line center can be attributed
to the anharmonicity in the vibrational potential, which leads to
the decay of optical phonons into low-energy acoustic phonons. The
effect of this decay on the frequency of the optical modes at $\varGamma$
has been described by the expression\citep{anand_96} 
\begin{equation}
\Delta\omega(T)=\omega(T)-\omega_{0}=C\left(1+\frac{2}{\exp\left[\hbar\omega_{0}/2k_{B}T\right]-1}\right)\:,\label{eq:fit_delta_w}
\end{equation}
where $\omega_{0}$ is the intrinsic frequency of the optical phonon,
i.e., $\omega_{0}\equiv\omega_{H}(T\rightarrow0)=1798$ cm$^{-1}$,
and $C$ is the anharmonic constant. The least-squares fit of the
HLR results to Eq. (\ref{eq:fit_delta_w}) is shown in Fig. \ref{fig:e2g_temp}.
The anharmonic constant amounts to $C=-64.7$ cm$^{-1}$. At 300 K
the temperature coefficient $\varphi_{T}=-0.026$ cm$^{-1}$/K derived
from the quantum simulation is about 5 times smaller that the classical
prediction. The quantum result is in excellent agreement with the
temperature coefficient derived from Raman measurements for single-walled
carbon nanotubes with a wide range of diameters,\citep{zhang_07}
although the diameter of the nanotube did not have any obvious influence
on the value of $\varphi_{T}.$ The experimental average value over
nanotubes of different diameters was $\varphi_{T}=-0.026$ cm$^{-1}$/K,
a value which was also obtained in subsequent Raman investigations.\citep{zhang_14}

\subsection{Temperature dependence of elastic moduli in graphene}

The temperature dependence of the elastic moduli is another important
anharmonic effect,\citep{katnelson_2005} that will be analyzed in
graphene by the HLR approach.

\subsubsection{Computational conditions}

Classical MD simulations of graphene in the isotropic $NPT$ ensemble
were performed with the LCBOPII model at discrete temperatures between
1 and 1000 K for an unstressed layer ($P=0$). The simulation cell
with $N=960$ atoms was the same as that one described in Subsec.
\ref{subsec:Graphene}. The time step was set as 1 fs. Two sets of
simulations were performed. In the first one, the carbon atoms move
unconstrained along trajectories with $2.4\times10^{7}$ MDS. In the
second set, the carbon atoms moved along trajectories with $8\times10^{6}$
MDS in the $(x,y)$-plane (the atomic $z$-coordinate is constraint
as $z=0$) . The acoustic LA and TA phonon dispersions 
(see Fig. \ref{fig:w_k_graphene})
were derived by a HLR analysis of the trajectories. The slopes of
the acoustic TA and LA phonons of graphene in the long-wavelength
limit ($k\rightarrow0$) (see Fig. \ref{fig:w_k_graphene}) correspond
to the sound velocities $v_{T}$ and $v_{L}$. They were determined
by a least-squares fit of the corresponding phonon dispersion bands
using the function
\begin{equation}
   \omega=\left(v^{2}k^{2}+ck^{4}\right)^{1/2}\:,
\end{equation}
where $v$ is the sound velocity and $c$ is a fitting constant. Only
those $\mathbf{k}$-points with modulus $k < 0.43$ \AA$^{-1}$
were included in the fit.

\subsubsection{Elastic moduli from the HLR analysis}

\begin{figure}
\vspace{0.1cm}
\includegraphics[width=7.0cm]{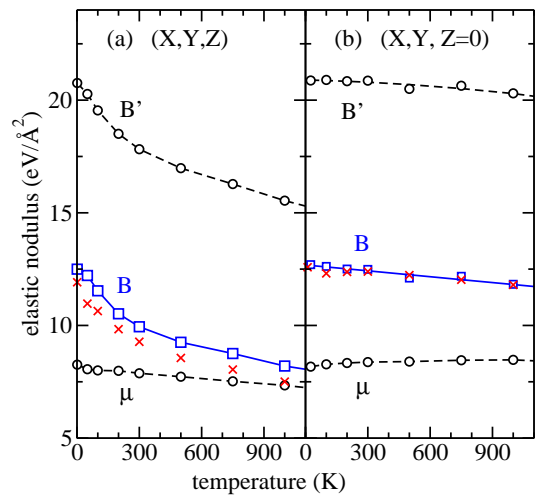}
\vspace{0.4cm}
\caption{Temperature dependence of elastic moduli of graphene as derived from
the HLR analysis of classical MD simulations. Open circles are results
for $\mu$ and $B'$ derived from a least-squares fit of the phonon
dispersion curves. Open squares are the estimation of the in-plane
stiffness as $B=B'-\mu$. The crosses are values of $B$ from the
fluctuation formula in Eq. (\ref{eq:B_fluctuation}). Lines are guides
to the eye. Panel (a) shows results of simulations where the layer
atoms move without constraints in the $(x,y,z)$ space. In Panel (b)
the carbon atoms move in the $(x,y)$-plane with the constraint $z=0$.
The results were derived with the LCBOPII model.}
\label{fig:elastic_constants}
\end{figure}

The sound velocities of the acoustic (TA and LA) phonons of 2D layers
provide information on its elastic moduli. The sound velocities of
the TA and LA branches for an isotropic 2D elastic media are\citep{cottam_15}
\begin{equation}
v_{T}=\left(\frac{\mu}{\rho}\right)^{1/2}\:,
\end{equation}
\begin{equation}
v_{L}=\left(\frac{B'}{\rho}\right)^{1/2}\:.
\end{equation}
where $\rho$ is the in-plane mass density, $\mu$ is the shear modulus
(Lamé's second coefficient), and $B'$ is the unilateral compressional
modulus defined as,\citep{behroozi_96}
\begin{equation}
       B'=\lambda+2\mu\:,
\end{equation}
where $\lambda$ is the Lamé's first coefficient.\citep{behroozi_96}
The in-plane compressional modulus (the 2D analogous to the bulk modulus)
is derived from $\mu$ and $B'$ as\citep{behroozi_96}
\begin{equation}
     B = B'-\mu=\lambda+\mu\:.\label{eq:B_lande}
\end{equation}
The temperature dependence of the elastic moduli, $\mu$ and $B'$
of graphene, as derived from the HLR analysis are presented in Fig.
\ref{fig:elastic_constants} as open circles. In the strictly flat
2D layer ($z=0$ for the C atoms, Fig. \ref{fig:elastic_constants}b)
the shear modulus $\mu$ is nearly temperature independent, while
the moduli $B'$ and $B$ decrease slightly with temperature. 

It is interesting to compare the HLR result for the compressional
modulus, $B$ (open squares in Fig. \ref{fig:elastic_constants}b),
to the value calculated by the fluctuation formula in the 
$NPT$ ensemble,\citep{landau_80,ramirez17}
\begin{equation}
   B=\frac{V}{N\beta\delta V^{2}}\:,\label{eq:B_fluctuation}
\end{equation}
where $V$ denotes the average in-plane area per atom that fluctuates
in the $NPT$ ensemble with dispersion $\delta V^{2}$. The values
of $B$ derived from this equation are presented by crosses in Fig.
\ref{fig:elastic_constants}b. Within the statistical error of the
simulations, the values of $B$ by the HLR analysis are in good agreement
with the fluctuation formula. 

The results for $B'$, $\mu$, and $B$ as derived by the HLR analysis
when the atoms of the layer move without constraint are displayed
in Fig. \ref{fig:elastic_constants}a. The atomic out-of-plane vibrations
have a significant influence in the elastic moduli of the layer, that
display a more pronounced temperature dependence than in 
Fig. \ref{fig:elastic_constants}b.
A striking result is that the HLR values of the in-plane compressional
modulus $B$ (open squares) are now systematically larger than those
derived from the fluctuation formula (crosses) in Fig. \ref{fig:elastic_constants}a.
The atomic displacements in the out-of-plane direction, causing the
spatial corrugation of the 2D layer, must be the origin of this unexpected
behavior.

Several plausible effects may explain the fact that Eqs. (\ref{eq:B_lande})
and (\ref{eq:B_fluctuation}) give different results for a 2D layer
fluctuating in 3 dimensions. One possibility is that the anharmonicity
of the out-of-plane displacements is not correctly reproduced by the
HLR analysis. However, any anharmonic effect in a classical simulation
should decrease gradually with temperature and eventually vanish in
the low temperature limit, a behavior that is not supported by the
results of Fig. \ref{fig:elastic_constants}a.

A second possibility is that the disagreement between the results
derived from Eqs. (\ref{eq:B_lande}) and (\ref{eq:B_fluctuation})
is a finite size effect related to the out-of-plane vibrations. We
have checked the finite size effect at 300 K by increasing the cell
size from $N=960$ to $N=8400$ atoms. The size effect reduces the
value of $B$. However, the decrease derived from the HLR phonon dispersion
relations is lower ($\sim21\%$) than that derived from the fluctuation
formula ($\sim27\%$). Thus, this effect would make the disagreement
found between both results even larger.

An intriguing explanation for the disagreement in the results of $B$
in Fig. \ref{fig:elastic_constants}a and the agreement in 
Fig. \ref{fig:elastic_constants}b,
is that when the 2D layer fluctuates in the $z$-direction, forming
ripples and loosing its strictly 2D character, the in-plane compressional
modulus $B$ in Eq. (\ref{eq:B_lande}) becomes different than that
one in Eq. (\ref{eq:B_fluctuation}). Ripples depend on an additional
elastic modulus, the bending stiffness, $\kappa$, that for an atom
thick layer is a constant independent from the Lamé coefficients,
$\lambda$ and $\mu$. 

\section{Applications: graphene bilayer and graphane
\label{sec:Applications:-graphene-bilayer}}

Examples of the application of the HLR approach to 2D layers different
from graphene monolayer is illustrated for a graphene bilayer and
for chair-graphane.

\subsection{Harmonic phonon dispersion in a graphene bilayer}

\begin{figure}
\vspace{0.1cm}
\includegraphics[width=7.0cm]{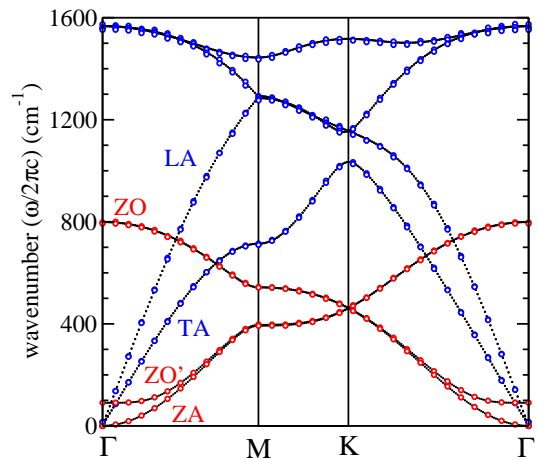}
\vspace{0.7cm}
\caption{The harmonic phonon dispersion of bilayer graphene with stacking AB,
as derived from the dynamical matrix of the LCBOPII model, is shown
by dotted lines that look like solid lines when the slope is small.
Open circles represent the HLR result as derived from a classical
$NPT$ simulation using the same model. The MD simulation was performed
close to the harmonic $T\rightarrow0$ limit, at 1 K and $P=0$, using
a supercell with $N=1920$ atoms. The $12\times12$ susceptibility
tensor $\chi(\mathbf{k})$ displays a block structure with separation
of $(X,Y)$ (blue circles) and $Z$ (red circles) bands. }
\label{fig: w_k_bilayer}
\end{figure}

A classical MD $NPT$ simulation of a graphene bilayer with AB stacking
has been performed at $T=1$ K and $P=0$ with similar computational
conditions as those employed for graphene in the previous Subsec.
\ref{subsec:Graphene}. The hexagonal cell parameter with the employed
LCBOPII model amounts to 2.4584 \AA, while the interlayer
distance is 3.3372 \AA. The simulation cell includes
two atomic layers adding up $N=1920$ atoms. The dimension of the
tensor $\chi(\mathbf{k})$ is $12\times12$ for the bilayer. A comparison
of the harmonic dispersion curves, $\omega_{H}(\mathbf{k})$, with
those derived by the HLR approach is presented in Fig. \ref{fig: w_k_bilayer}.
We find that the agreement between both methods is excellent for the
12 vibrational bands. 

All vibrational bands of bilayer graphene, except ZA (see Fig. \ref{fig: w_k_bilayer}),
appear as degenerate at the scale of the figure. The ZA band of graphene
is splitted into ZA and ZO' bands in the bilayer. The larger splitting
is found at $\varGamma$ where the frequency of the ZO' mode amounts
to 91 cm$^{-1}$ (see Fig. \ref{fig: w_k_bilayer}). This is the layer
breathing mode, where the $z$-distance between the two graphene layers
oscillates around its equilibrium value. The acoustic LA and TA modes
of graphene also split in the bilayer, but the splitting is too small
to be seen in Fig. \ref{fig: w_k_bilayer}. The largest splitting
is again found at $\varGamma$ and amounts to 15 cm$^{-1}$. This
vibrational mode of the bilayer corresponds to a rigid layer shear
mode, involving the relative motion of atoms in adjacent planes. Raman
spectroscopy of a graphene bilayer gives a frequency of 32 cm$^{-1}$
for this rigid layer shear mode,\citep{tan_12} while the ZO' breathing
mode appears at 89 cm$^{-1}$ at $\varGamma$.\citep{lin_18} The
splitting of the LO and TO modes at $\varGamma$ is even smaller,
less that 1 cm$^{-1}$ with the LCBOPII model. The splitting obtained
with density-functional perturbation theory amounts to 5 cm$^{-1}$.\citep{yan_08}

\subsection{Vibrational anharmonic effects in chair-graphane\label{subsec:Vibrational-anharmonic-effects}}

\subsubsection{Computational conditions}

\begin{figure}
\vspace{0.1cm}
\includegraphics[width=7.0cm]{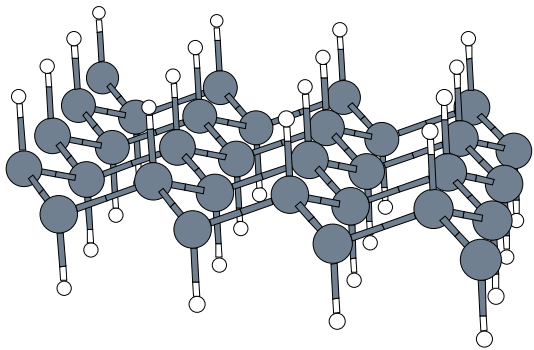}
\vspace{0.4cm}
\caption{2D structure of chair-graphane. Carbon and hydrogen atoms are represented
by circles, the smaller ones correspond to H.}
\label{fig:graphane}
\end{figure}

The 2D layer structure of chair-graphane is displayed in Fig. \ref{fig:graphane}.
The electronic structure has been treated with the TB electronic band
structure method of Subsec. \ref{subsec:Temperature-dependence-of}.
The hexagonal cell displays a cell parameter of 2.5320 \AA,
and the vertical atomic distances to the mean layer plane are 0.2319 \AA\
for C and 1.3576 \AA\ for H. Classical MD simulations
with the isotropic $NPT$ ensemble were performed on a $6\times4$
supercell of a centered rectangular cell containing $N=192$ atoms,
at external stress $P=0$, and temperatures of $T=1$ K and $T=300$
K. Only the $\varGamma$ point for the sampling of the BZ of the simulation
supercell was used in the electronic structure calculation. The equilibration
run consisted on $2\times10^{5}$ MDS and a trajectory with 10000
layer configurations was stored at equidistant steps from a run with
$2\times10^{6}$ MDS. The classical MD simulation was performed with
a time step of 1 fs.

\subsubsection{Phonon dispersion}

\begin{figure}
\vspace{0.1cm}
\includegraphics[width=7.0cm]{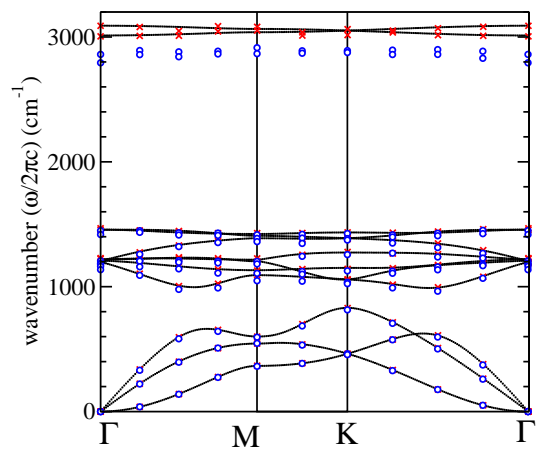}
\vspace{0.8cm}
\caption{The harmonic phonon dispersion of chair-graphane, as calculated from
the dynamical matrix of the employed TB model, is shown by lines.
HLR wavenumbers derived from classical $NPT$ simulations with the
same TB model at $P=0$, using a supercell with $N=192$ atoms, are
shown at temperatures of 300 K (open circles) and 1 K (crosses). A
significant anharmonic red-shift is visible in the bands associated
to the C$-$H stretching vibrations at frequencies around 2850 cm$^{-1}.$
The HLR results for the other vibrational bands at 300 K and at 1K
are nearly indistinguishable at the scale of the figure.}
\label{fig:w_k_graphane_h}
\end{figure}

The harmonic phonon dispersion of chair-graphane, as derived from
the dynamical matrix of the employed TB model, is displayed in Fig.
\ref{fig:w_k_graphane_h} (lines). There appear 12 dispersion bands.
The two flat dispersion bands around 3000 cm$^{-1}$ correspond to
the stretching of the C$-$H bonds along the $z$-direction perpendicular
to the layer. The general appearance of the TB harmonic dispersion
bands show reasonable agreement to previous calculations by first-principles
electronic structure calculations.\citep{cadelano_10,peelaers_2011}
The result of the HLR analysis of a classical graphane simulation
at 1 K, displayed in Fig. \ref{fig:w_k_graphane_h} (crosses), is
identical, within the statistical error of the simulation, to the
harmonic dispersion bands.

Anharmonic effects due to the increased amplitude of atomic vibrations
are expected to appear as the temperature rises. Therefore, at 300
K deviations of the HLR phonon dispersion from the harmonic limit
indicate the presence of anharmonic effects in the corresponding vibrational
modes. The HLR phonon dispersion curves at 300 K is displayed by open
circles in Fig. \ref{fig:w_k_graphane_h}. The most significant anharmonic
effect is a red-shift of the bands associated to the stretching of
the C$-$H bonds by about 150 cm$^{-1}$, while anharmonic effects
in the rest of vibrational bands are comparatively low. For many organic
molecules, the anharmonicity of C\textendash H modes is estimated
to red-shift the harmonic stretching frequency by about 120\textendash 140
cm$^{-1}$.\citep{gentile_2017} 

\section{Summary}

We have presented the HLR method as a tool to study the phonon dispersion
relations of 2D layers from the analysis of trajectories generated
by equilibrium MD or MC simulations, either with the quantum PI formalism,
or in the classical limit. The HLR method is based on the analysis
of the spatial correlations of centroid or atomic displacements by
means of the diagonalization of the covariance matrices associated
to these displacements. The HLR method for 2D layers could be straightforwardly
applied to 1D or 3D solids. The only difference is that in the Fourier
transform of Eq. (\ref{eq:FT}) the dimension of the $\mathbf{r}_{eq}$-
and $\mathbf{k}$-vectors will depend on the the dimensionality of
the problem. 

The physical information needed in the HLR method is the static susceptibility
tensor, $\chi(\mathbf{k})$, that defines the linear response of an
equilibrium system to static forces. As the linear response of the
system depends on the anharmonicity of the interatomic potential,
anharmonic effects are included in the HLR approach in a realistic
way.

By means of classical MD simulations at very low temperatures, we
have checked that the HLR approach reproduces the harmonic phonon
dispersion relations obtained by diagonalization of the dynamical
matrix. This test has been presented for several 2D systems: a graphene
monolayer, a graphene bilayer, and a graphane monolayer. The sensitivity
of the HLR approach to reproduce anharmonic effects has been demonstrated
by the calculation of the temperature dependence of the atomic kinetic
energy of carbon atoms in graphene by quantum PIMD simulations. Anharmonic
shifts in the optical phonon frequencies of graphene have been derived
by the HLR method. The comparison of the quantum results and the classical
limit display the significant anharmonicity of the zero-point vibrations.
Another studied anharmonic effect is the temperature dependence of
the elastic moduli of graphene. An intriguing difference has been
found in alternative ways to calculate the in-plane compressional
modulus of a graphene layer, either by the HLR phonon dispersion curves
or by the fluctuation of the area in the isothermal-isobaric ensemble.
Both methods provide identical result for a strict 2D planar layer,
but display a systematic difference in the presence of out-of-plane
fluctuations of the layer.

\acknowledgments 

This work was supported by Dirección General de Investigación, MINECO
(Spain) through Grants No. FIS2015-64222-C2-1-P and PGC2018-096955-B-C44. 

\bibliographystyle{apsrev4-1}
\newpage{}

\end{document}